\documentclass{PoS}

\usepackage{amstext}  
\usepackage{amsfonts}
\usepackage{amssymb} 
\usepackage{amsmath}
\usepackage{amsthm}
\usepackage{xcolor} 
\usepackage{color}
\usepackage[hyphens]{url} 

\usepackage{fancyhdr}
\usepackage{cite}
\usepackage{slashed}
\usepackage{pstricks}
\usepackage{axodraw4j}
\usepackage{graphicx}
\usepackage{multirow}

\newcommand{\p}{\partial}

\newcommand{\f}[2]{\frac{#1}{#2}}
\newcommand{\sss}[1]{\scriptscriptstyle#1}
\newcommand{\ssst}[1]{\scriptscriptstyle{\text{#1}}}
\newcommand{\vv}[2]{\left( \begin{array}{c} #1 \\ #2  \end{array} \right)}
\newcommand{\bea}{\begin{eqnarray}}
\newcommand{\eea}{\end{eqnarray}}
\newcommand{\be}{\begin{equation}}
\newcommand{\ee}{\end{equation}}
\newcommand{\ba}{\begin{align}}
\newcommand{\ea}{\end{align}}
\newcommand{\beas}{\begin{eqnarray*}}
\newcommand{\eeas}{\end{eqnarray*}}
\newcommand{\bes}{\begin{equation*}}
\newcommand{\ees}{\end{equation*}}
\newcommand{\bas}{\begin{align*}}
\newcommand{\eas}{\end{align*}}

\newcommand{\eps}{{\varepsilon}}
\newcommand{\cd}{{\cdot}}

\newcommand{\gs}{g_{\scriptscriptstyle{s}}}
\newcommand{\yt}{y_{\scriptscriptstyle{t}}}

\newcommand{\gb}{g_1}
\newcommand{\gw}{g_2}
\newcommand{\gaf}{\gamma_{\scriptscriptstyle{5}}}
\newcommand{\als}{\alpha_{\scriptscriptstyle{s}}}

\newcommand{\lb}{\left(}
\newcommand{\rb}{\right)}

\newcommand{\msbar}{$\overline{\text{MS}}$}

\definecolor{bluemar}{rgb}{0,0,.5}
\definecolor{redmar}{rgb}{.8,0,0}
\definecolor{greenmar}{rgb}{0,.5,0}

\title{Three-loop beta function for the Higgs self-coupling}
\ShortTitle{Three-loop beta function for the Higgs self-coupling}

\author{\speaker{Max Zoller}          \thanks{Report number: TTP14-021, SFB/CPP-14-40}\\
        Karlsruhe Institute of Technology (KIT)\\
        E-mail: \email{max.zoller@kit.edu}}

\abstract{
In the last two years the renormalization group functions for the couplings and fields of the Standard Model have been computed
at three-loop level \cite{PhysRevLett.108.151602,Mihaila:2012pz,Chetyrkin:2012rz,Chetyrkin:2013wya,Bednyakov:2012rb,Bednyakov:2013eba,Bednyakov:2012en}.
The evolution of the self-coupling $\lambda$ of the Standard Model Higgs boson is of particular importance 
due to its close connection with the stability of the Standard Model vacuum state.
In this talk the three-loop corrections to the $\beta$-function for this crucial coupling are discussed.

The calculation of three-loop $\beta$-functions and anomalous dimensions poses special technical challenges,
such as the huge number of diagrams and the proper treatment of $\gaf$ in dimensional regularization.
In order to avoid infrared divergences resulting from setting external momenta to zero in the case of 
the Higgs self-coupling an auxiliary mass is used to compute the ultraviolet divergences needed for the renormalization constants.
This method, first suggested in \cite{Misiak:1994zw}, is explained in some detail.

Finally, an update for the status of the vacuum stability problem in the Standard Model up to the Planck scale is presented.
}

\FullConference{Loops and Legs in Quantum Field Theory\\
		27 April 2014 - 02 May 2014\\
		Weimar, Germany}

\begin{document}

\section{Motivation: The vacuum stability problem} 

The Standard Model of particle physics describes the interactions of fermions through the exchange of gauge bosons.
In the covariant derivative 
\be
D^\mu=\p^\mu-i \gb Y B^\mu-i \f{\gw}{2} \sigma^a W^{a\,\mu}-i\gs T^a A^{a\,\mu}{}. \label{covderiv}
\ee
the three gauge couplings $\gs$, $\gw$ and $\gb$ of the $SU(3)\times SU(2) \times U(1)$ group are defined.\footnote{$B^\mu$, $W^{a\,\mu}$ and $A^{a\,\mu}$ are the gauge fields
of $SU(3)$, $SU(2)$ and $U(1)$ respectively, $\sigma^a$ are the Pauli matrices, $T^a$ are the generators of $SU(3)$ and $Y$ the $U(1)$ hypercharge of the field on which $D^\mu$ acts.}
In addition to the fermions and gauge bosons a scalar $SU(2)$ doublet is introduced which aquires a vacuum expectation value (VEV)
at the electroweak scale $v\approx 246.2$ GeV.
The fermion masses and the Higgs-fermion interaction are derived from the Yukawa sector of the Standard Model and
the Higgs self-interaction is introduced to the Lagrangian in the classical Higgs potential
\be V(|\Phi|)=\left(m^2\, \Phi^\dagger\Phi+\lambda\, \lb\Phi^\dagger\Phi\rb^2\right), \quad \Phi=\vv{\Phi_1}{\Phi_2} {}.\ee
After spontaneous symmetry breaking we have the classical field strength
\be \Phi_{\ssst{cl}}:=\langle 0 |\Phi|0\rangle=\f{1}{\sqrt{2}}\vv{0}{v} \ee
and describe excitations from the ground state with four quantum fields, the Higgs field $H$ and three Goldstone bosons $\chi$, $\Phi^{\pm}$:
\be \Phi=\vv{\Phi_1}{\Phi_2}\underrightarrow{\text{SSB}} \vv{\Phi^+}{\f{1}{\sqrt{2}}(v+H-i\chi)}. \ee

\begin{figure}[!b]
\begin{tabular}{ccc}
\begin{picture}(130,90)(0,0)
\put(0,0){\includegraphics[width=0.31\linewidth]{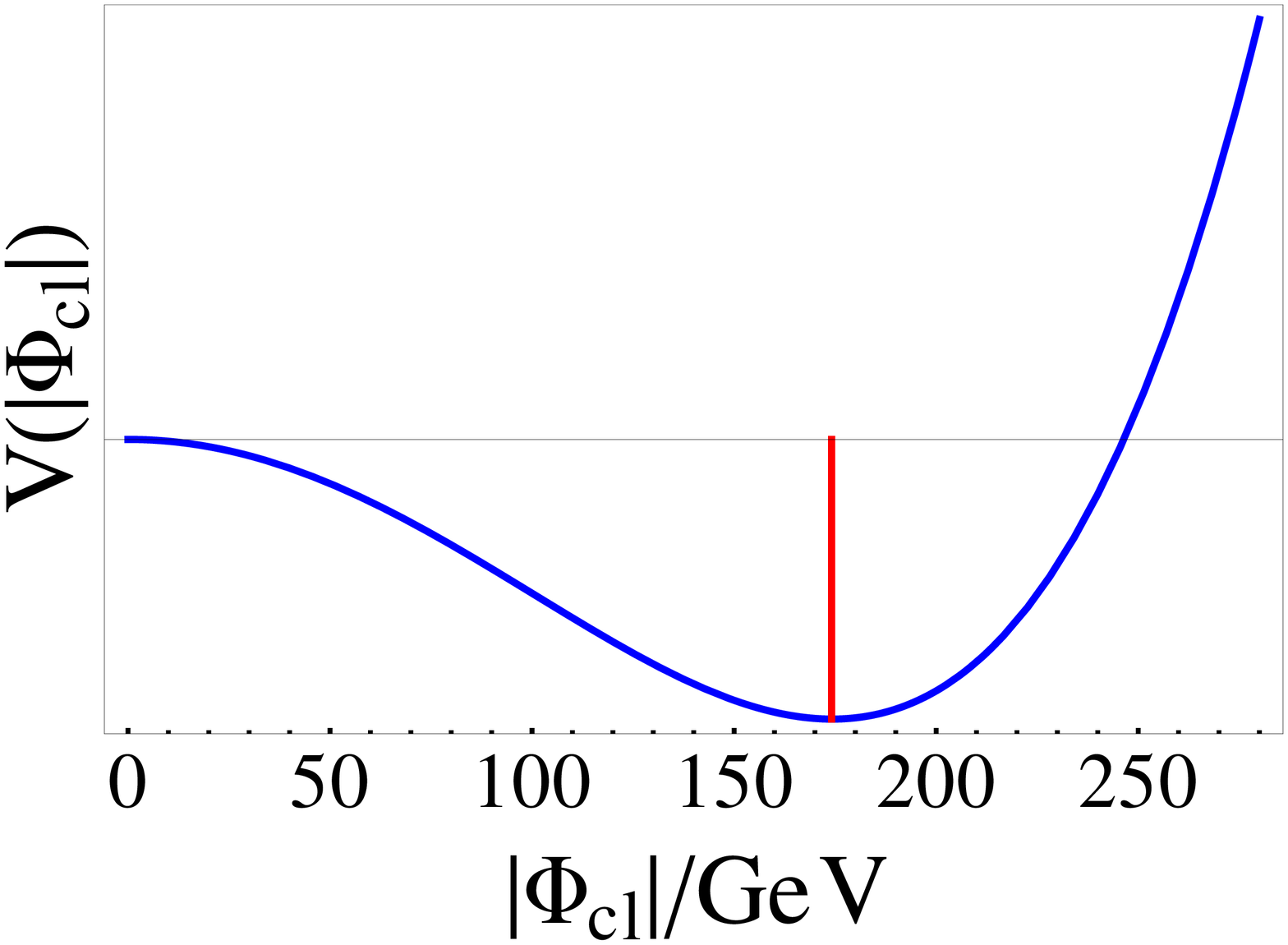}}
\Text(79,57)[cc]{{\Red{$\f{v}{\sqrt{2}}$}}}
 \end{picture}
&
\begin{picture}(135,90)(0,0)
\put(0,0){\includegraphics[width=0.345\linewidth]{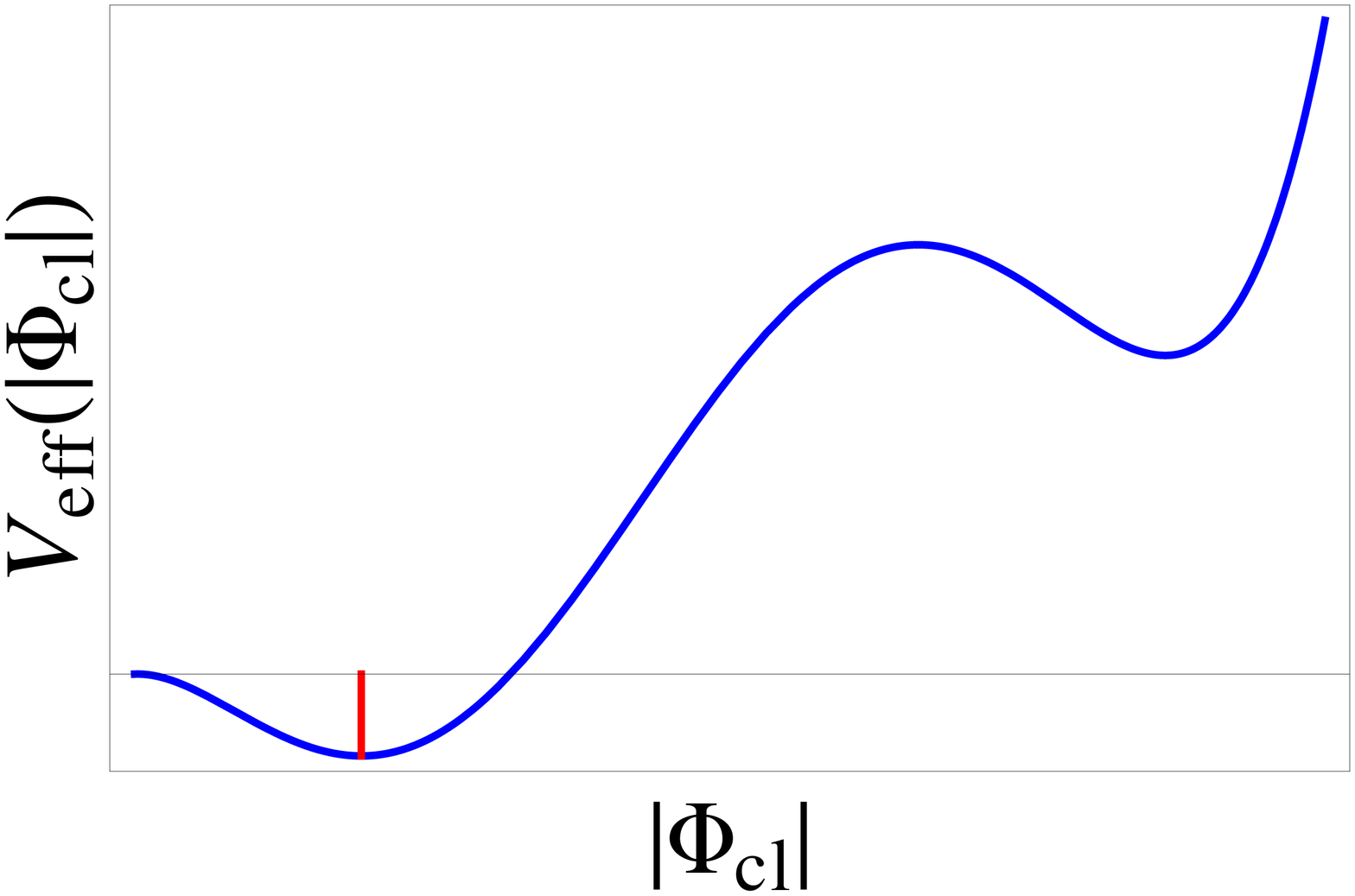}}
\Text(35,30)[cc]{{\Red{$\f{v}{\sqrt{2}}$}}}
 \end{picture}
&
\begin{picture}(130,90)(0,0)
\put(0,0){\includegraphics[width=0.345\linewidth]{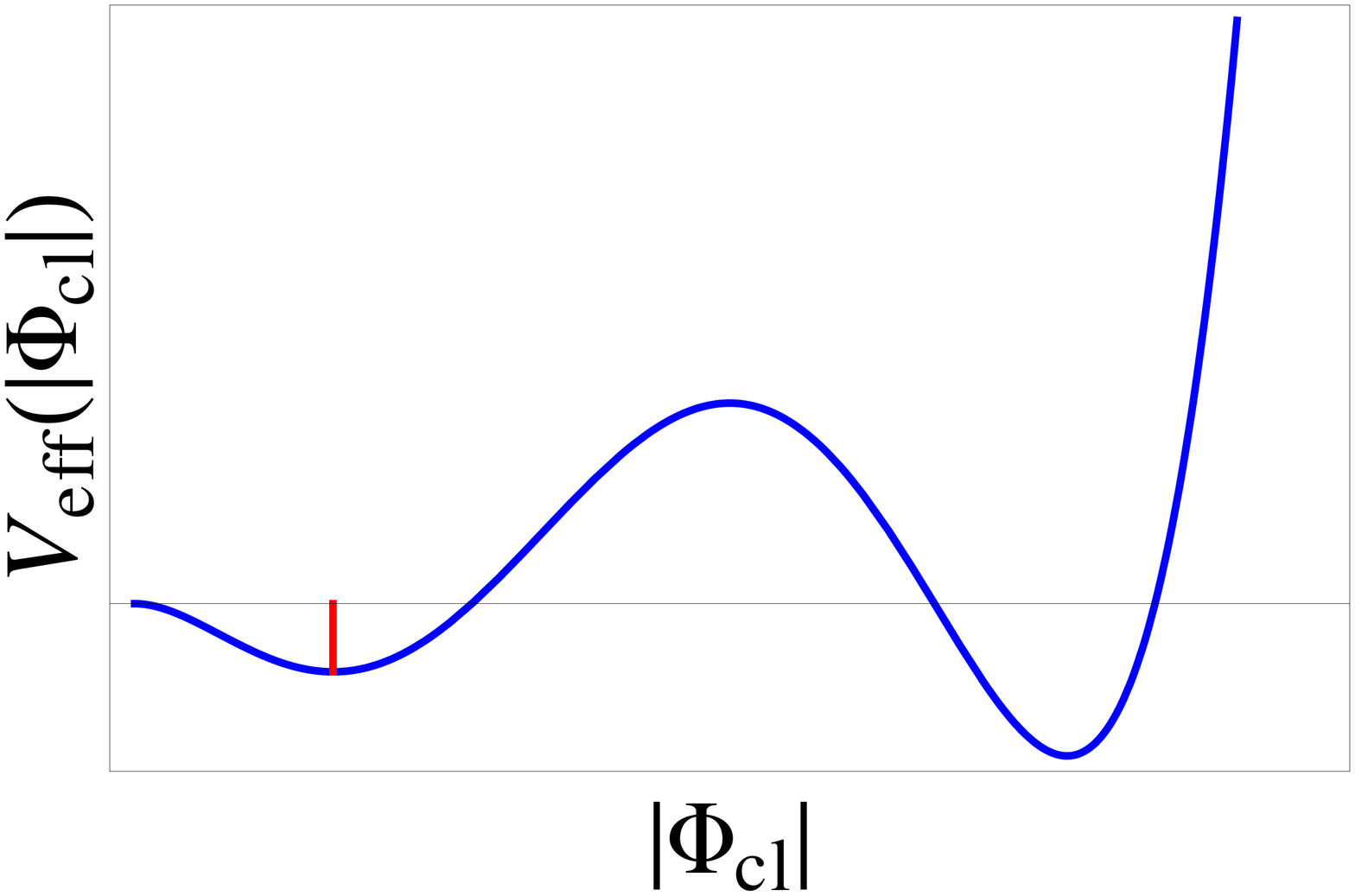}}
\Text(34,38)[cc]{{\Red{$\f{v}{\sqrt{2}}$}}}
 \end{picture}\\
classical Higgs potential & stable & unstable/metastable
\end{tabular}
\caption{Classical and effective Higgs potential} \label{Veffective} 
\end{figure}

Due to radiative corrections the effective couplings evolve with the renormalization scale and we have to consider the effective Higgs potential \cite{PhysRevD.7.1888} 
instead of the classical one. The effective potential develops a second minimum at high classical field strengths $|\Phi_{\ssst{cl}}|$ if we extrapolate the Standard Model up to e.g.~the Planck scale
\mbox{$M_{\ssst{Planck}}\sim 10^{19}$ GeV.} In the absence of physics beyond the Standard Model at the LHC so far this is a conceivable scenario.
The effective potential is a function of $\Phi_{\ssst{cl}}$ and is affected by the self-interactions 
of the scalar fields as well as the interactions of the scalar fields with all other fields. Hence it depends on all couplings of the theory.

Generic shapes of the classical Higgs potential and of the effective potential are shown in Fig.~\ref{Veffective} for the cases
of a Higgs mass larger and smaller than a critical value $m_{\ssst{min}}$, the minimal stability bound.
For \mbox{$M_{\ssst{H}}>m_{\ssst{min}}$} the second minimum is higher than the one at the electroweak scale and therefore
the vacuum state with \mbox{$|\Phi_{\ssst{cl}}|=\f{v}{\sqrt{2}}$} is stable. However, for \mbox{$M_{\ssst{H}}<m_{\ssst{min}}$} 
the second minimum is energetically favoured and the electroweak vacuum state is not 
stable against decay into this global ground state. Depending on whether the lifetime of the electroweak vacuum state 
is shorter or longer than the age of the universe this is called an unstable or metastable
scenario.
It has been demonstrated that the stability of the Standard Model vacuum is in good approximation
equivalent to the question whether the Higgs self-interaction $\lambda$ stays positive up to the maximum validity scale of the theory, e.g. $M_{\ssst{Planck}}$
\cite{Altarelli1994141,Cabibbo:1979ay,Ford:1992mv}. A detailed analysis of the vacuum stability problem in the Standard Model can be found in
\cite{Buttazzo:2013uya,Masina:2012tz,Bezrukov:2012sa,Degrassi:2012ry,Zoller:2012cv,Chetyrkin:2012rz,EliasMiro:2011aa,Holthausen:2011aa,Zoller:2013mra}.

The evolution of any coupling is described by the respective $\beta$-function
\be \beta_{X}(\lambda,\yt,\gs,g_2,g_1)=\mu^2\f{d}{d\mu^2} X(\mu),\quad X \in \left\{\lambda,\yt,\gs,g_2,g_1\right\}. \label{betasystem} \ee
Every $\beta$-function is a power series in all couplings of the theory which is why the $\beta$-functions for all numerically relevant couplings
are needed. Recently, the $\beta$-functions for the gauge couplings \cite{PhysRevLett.108.151602,Mihaila:2012pz,Bednyakov:2012rb}, Yukawa couplings \cite{Chetyrkin:2012rz,Bednyakov:2012en}
and for the Higgs self-interaction \cite{Chetyrkin:2012rz,Chetyrkin:2013wya,Bednyakov:2013eba} have been computed at three-loop level.

In order to determine the evolution of $\lambda$ the coupled system of differential equations \eqref{betasystem} needs to be solved
using initial conditions for all couplings, e.g. their value at the scale of the top pole mass.
In table \ref{SMKopplungenWertewitherrors} the values for the numerically largest couplings are given at this scale. These are derived by matching
the experimentally accessible parameters $G_{\ssst{F}}$, $M_{\ssst{t}}$, $M_{\ssst{H}}$, $M_{\ssst{W}}$, $M_{\ssst{Z}}$ and $\als^{\text{\msbar}}(M_{\ssst{Z}})$ to the \msbar-parameters
$\gs(M_{\sss{t}})$, $\gw(M_{\sss{t}})$, $\gb(M_{\sss{t}})$, $\yt(M_{\sss{t}})$ and $\lambda(M_{\sss{t}})$.
One-loop \cite{Tarrach:1980up,Sirlin1986389,Hempfling:1994ar} and two-loop corrections \cite{Gray:1990yh,Jegerlehner:2001fb,Jegerlehner:2002em,Faisst:2003px,
Eiras:2005yt,Jegerlehner:2003py,Bezrukov:2012sa,Buttazzo:2013uya,Kniehl:2014yia} to this matching are taken into account.
Except for $\yt$ the Yukawa interactions $y_b,\,y_c,$ etc.~can be neglected in this context due to their smallness. The same
applies to the off-diagonal entries of the Yukawa matrices.

\begin{table}[!b] 
\begin{center}
\begin{tabular}{|c|c|}\hline
coupling & value for $\mu=M_{\sss{t}}$\tabularnewline \hline
 $\gs$ &  $1.1666$\tabularnewline
 $\gw$ & $0.6483$\tabularnewline 
 $\gb$ & $0.3587$\tabularnewline 
 $\lambda$ & $0.1276$\tabularnewline
 $\yt$ & $0.9369$\tabularnewline
\hline
\end{tabular} 
\caption{Standard Model couplings at the top pole mass scale for \mbox{$M_{\sss{t}}=173.34$ GeV \cite{ATLAS:2014wva},} 
\mbox{$M_{\sss{H}}=125.9$ GeV \cite{ATLAS:2012ae,Chatrchyan:2012tx,TEVNPH:2012ab}} und 
\mbox{$\als(M_{\sss{Z}})=0.1184$  \cite{Bethke:2012jm}} using the on-shell to \msbar-relations given in \cite{Buttazzo:2013uya}.}
\end{center}
\label{SMKopplungenWertewitherrors}
\end{table}

\section{Calculating the $\beta$-function for the Higgs self-interaction with an auxiliary mass}

The $\beta$-function of a coupling is computed in dimensional regularization from the renormalization constant of a vertex proportional to this coupling
and the renormalization constants for the external fields of this vertex. The calculation is performed in the unbroken phase of the Standard Model and the results are given in the \msbar-scheme.
There are many challenges to the computation of the three-loop $\beta$-function for the Higgs self-interaction. One is the huge number of Feynman diagrams,
573692 for the $\Phi_1^4$-vertex even if we factorize the full gauge group factor from the momentum space factor for each diagram (for details see \cite{Chetyrkin:2013wya}).
Another issue is the proper treatment of $\gaf$ in dimensional regularization. 
Whereas a naive treatment of $\gaf$ does not work for the computation of $\beta_{\yt}$ we have shown in \cite{Chetyrkin:2012rz,Chetyrkin:2013wya}
that a naive treatment is sufficient for $\beta_{\lambda}$ at three-loop level.

\begin{figure}[!tb]
\begin{center}\scalebox{0.9}{
$\begin{picture}(50,55) (0,0)
\DashArrowLine(0,25)(25,0){3}
\DashArrowLine(25,0)(50,25){3}
\DashArrowLine(25,0)(0,-25){3}
\DashArrowLine(50,-25)(25,0){3}
\CCirc(25,0){10}{Black}{Blue}
\SetColor{Red}
\LongArrow(-5,20)(5,10)
\LongArrow(55,20)(45,10)
\Text(-7,10)[cc]{\large{\Red{$0$}}}
\Text(57,10)[cc]{\large{\Red{$0$}}}
\LongArrowArcn(25,-38)(20,160,20)
\Text(25,-28)[cc]{\Large{\Red{$q$}}}
\SetColor{Black}
\CCirc(-8,33){7}{Black}{White}
\Text(-8,33)[cc]{\large{\Black{$3$}}}
\CCirc(58,33){7}{Black}{White}
\Text(58,33)[cc]{\large{\Black{$4$}}}
\CCirc(-8,-33){7}{Black}{White}
\Text(-8,-33)[cc]{\large{\Black{$1$}}}
\CCirc(58,-33){7}{Black}{White}
\Text(58,-33)[cc]{\large{\Black{$2$}}}
  \end{picture}
\qquad=\qquad
\begin{picture}(50,55) (0,0)
\DashArrowLine(0,25)(15,0){3}
\DashArrowLine(35,0)(50,25){3}
\DashArrowLine(15,0)(0,-25){3}
\DashArrowLine(50,-25)(35,0){3}
\DashArrowArc(25,0)(10,180,0){3}
\DashArrowArc(25,0)(10,0,180){3}
\Vertex(15,0){1}
\Vertex(35,0){1}
\SetColor{Red}
\LongArrowArcn(25,-38)(20,160,20)
\Text(25,-28)[cc]{\Large{\Red{$q$}}}
\SetColor{Black}
\CCirc(-8,33){7}{Black}{White}
\Text(-8,33)[cc]{\large{\Black{$3$}}}
\CCirc(58,33){7}{Black}{White}
\Text(58,33)[cc]{\large{\Black{$4$}}}
\CCirc(-8,-33){7}{Black}{White}
\Text(-8,-33)[cc]{\large{\Black{$1$}}}
\CCirc(58,-33){7}{Black}{White}
\Text(58,-33)[cc]{\large{\Black{$2$}}}
\Text(25,-55)[cc]{\large{\Black{(a)}}}
  \end{picture}
\qquad+\qquad
\begin{picture}(50,55) (0,0)
\DashArrowLine(15,0)(0,25){3}
\DashArrowLine(50,25)(35,0){3}
\DashArrowLine(15,0)(0,-25){3}
\DashArrowLine(50,-25)(35,0){3}
\DashArrowArcn(25,0)(10,0,180){3}
\DashArrowArc(25,0)(10,0,180){3}
\Vertex(15,0){1}
\Vertex(35,0){1}
\SetColor{Red}
\LongArrowArcn(25,-38)(20,160,20)
\Text(25,-28)[cc]{\Large{\Red{$q$}}}
\SetColor{Black}
\CCirc(-8,33){7}{Black}{White}
\Text(-8,33)[cc]{\large{\Black{$4$}}}
\CCirc(58,33){7}{Black}{White}
\Text(58,33)[cc]{\large{\Black{$3$}}}
\CCirc(-8,-33){7}{Black}{White}
\Text(-8,-33)[cc]{\large{\Black{$1$}}}
\CCirc(58,-33){7}{Black}{White}
\Text(58,-33)[cc]{\large{\Black{$2$}}}
\Text(25,-55)[cc]{\large{\Black{(b)}}}
  \end{picture}
\qquad+\qquad
\begin{picture}(50,55) (0,0)
\DashArrowLine(0,25)(25,10){3}
\DashArrowLine(25,10)(50,25){3}
\DashArrowLine(25,-10)(0,-25){3}
\DashArrowLine(50,-25)(25,-10){3}
\DashArrowArc(25,0)(10,90,270){3}
\DashArrowArc(25,0)(10,270,90){3}
\Vertex(25,10){1}
\Vertex(25,-10){1}
\SetColor{Red}
\LongArrowArcn(25,-38)(20,160,20)
\Text(25,-28)[cc]{\Large{\Red{$q$}}}
\SetColor{Black}
\CCirc(-8,33){7}{Black}{White}
\Text(-8,33)[cc]{\large{\Black{$3$}}}
\CCirc(58,33){7}{Black}{White}
\Text(58,33)[cc]{\large{\Black{$4$}}}
\CCirc(-8,-33){7}{Black}{White}
\Text(-8,-33)[cc]{\large{\Black{$1$}}}
\CCirc(58,-33){7}{Black}{White}
\Text(58,-33)[cc]{\large{\Black{$2$}}}
\Text(25,-55)[cc]{\large{\Black{(c)}}}
  \end{picture} \quad+\ldots$}\\[1.8cm]
  \end{center}
  \caption{One-loop diagrams contributing to the renormalization of the $\Phi_1^4$-vertex: An IR singularity appears in (c) if two external momenta are set to zero.}
    \label{4phiinfraredproblem}
\end{figure}
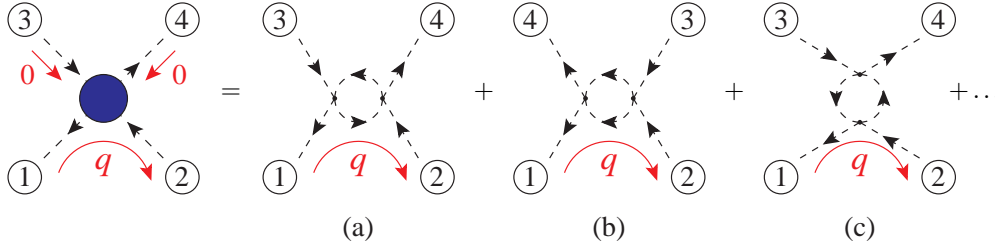

In the case of corrections to the quartic scalar vertex a problem arises already at one-loop level if we compute the renormalization constant setting two external momenta to zero
and then evaluating propagator-like massless diagrams. In Fig.~\ref{4phiinfraredproblem} we see that diagram (c) has an UV divergence as well as an IR one which cancel to give zero for the whole diagram.
In order to avoid the IR singularity and to retrieve the correct contribution to the UV renormalization constant we introduce an auxiliary mass in the denominator of every propagator.
Then we Taylor expand in the external momenta as far as needed in order to factorize the kinematic structure of the Greens function which we want to compute, e.g. $q^2 g^{\mu\nu}-q^\mu q^\nu$ for the gluon 
propagator. Now we can set the external momenta in the scalar part of the Greens function to zero because the UV divergence does not depend on those.

This method was suggested in \cite{Misiak:1994zw} and further developed in \cite{beta_den_comp}. In order to see why this method works let us consider the following decomposition
of a propagator denominator \cite{beta_den_comp}:
\be
\f{1}{(l+q)^2}=\f{1}{l^2-M^2}+\f{-q^2-2l\cd q-M^2}{l^2-M^2}\f{1}{(l+q)^2} \label{propdenomdecomp}
{},
\ee
where $l$ is a linear combination of loop momenta and $q$ of external momenta.
Using this formula recursively leads to the last term, which has $q$ in the denominator, contributing only to the finite part of the integral because
the power of the denominator is increased with every recursion. We consider now the case where two iterations are enough:
\be 
\begin{split}
\f{1}{(l+q)^2}=& \f{1}{l^2-M^2}+
\f{-q^2-2l\cd q}{(l^2-M^2)^2}
+\f{(-q^2-2l\cd q)^2}{(l^2-M^2)^3}\\
&-\f{M^2}{(l^2-M^2)^2}
+\f{M^2(M^2+2q^2+4l \cd q)}{(l^2-M^2)^3}   +
\f{(-q^2-2l\cd q-M^2)^3}{(l^2-M^2)^3}\f{1}{(l+q)^2}.
\end{split} \label{propdenomdecomprekursiv}
\ee
This decomposition is exact and hence using the left side of \eqref{propdenomdecomprekursiv} for every propagator denominator would give the correct and
$M^2$-independent result. As we are only interested in the UV divergent part of the integral we neglect the last term in \eqref{propdenomdecomprekursiv}.
Furthermore, we notice that the first line of \eqref{propdenomdecomprekursiv} is exactly the Taylor expansion in the external momenta mentioned above.

If we do not use this exact decomposition but simply put $M^2$ in every propagator denominator and expand in the external momenta 
we neglect exactly the terms \mbox{$\propto M^2$} in the second line of \eqref{propdenomdecomprekursiv}.
At one-loop level this is not a problem because we know that the exact result does not depend on $M^2$. We can therefore reconstruct the contribution
from the terms \mbox{$\propto M^2$} in \eqref{propdenomdecomprekursiv} as counterterms to the $M^2$-terms in our result.

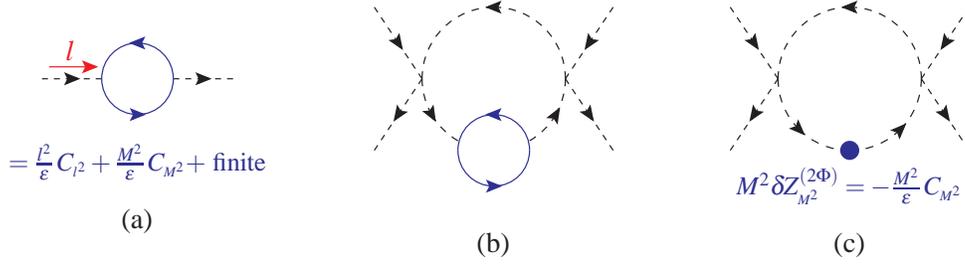
\begin{figure}[!tb]
\begin{center}\scalebox{0.9}{
\begin{tabular}{ccc}\,
\parbox{0.32\textwidth}{
\begin{picture}(80,40) (-30,0)
\DashArrowLine(0,0)(25,0){3}
\DashArrowLine(55,0)(80,0){3}
\SetColor{Blue}
\ArrowArc(40,0)(15,0,180)
\ArrowArc(40,0)(15,180,0)
\SetColor{Red}
\Text(12,12)[cc]{\large{\Red{$l$}}}
\LongArrow(3,5)(20,5)
\SetColor{Black}
\Text(40,-35)[cc]{{\Blue{$=\f{l^2}{\eps}\,C_{\sss{l^2}}+\f{M^2}{\eps}\,C_{\sss{M^2}}+$ finite}}}
\Text(40,-60)[cc]{\large{\Black{(a)}}}
  \end{picture}\\[2.3cm]} &
  \parbox{0.32\textwidth}{
\begin{picture}(80,40) (-30,0)
\DashArrowLine(-10,30)(10,0){3}
\DashArrowLine(10,0)(-10,-30){3}
\DashArrowLine(90,30)(70,0){3}
\DashArrowLine(70,0)(90,-30){3}
\DashArrowArc(40,0)(30,0,180){3}
\SetColor{Blue}
\ArrowArc(40,-30)(15,0,180)
\ArrowArc(40,-30)(15,180,0)
\SetColor{Black}
\DashArrowArc(40,0)(30,180,240){3}
\DashArrowArc(40,0)(30,300,0){3}
\Text(40,-70)[cc]{\large{\Black{(b)}}}
  \end{picture}\\[2.3cm]} & 
    \parbox{0.32\textwidth}{
\begin{picture}(80,40) (-30,0)
\DashArrowLine(-10,30)(10,0){3}
\DashArrowLine(10,0)(-10,-30){3}
\DashArrowLine(90,30)(70,0){3}
\DashArrowLine(70,0)(90,-30){3}
\DashArrowArc(40,0)(30,0,180){3}
\DashArrowArc(40,0)(30,180,270){3}
\DashArrowArc(40,0)(30,270,0){3}
\SetColor{Blue}
\Vertex(40,-30){4}
\Text(40,-45)[cc]{{\Blue{$M^2 \delta\!Z_{\sss{M^2}}^{(2\Phi)}=-\f{M^2}{\eps}\,C_{\sss{M^2}}$}}}
\SetColor{Black}
\Text(40,-70)[cc]{\large{\Black{(c)}}}
  \end{picture}\\[2.3cm]}
  \end{tabular}}
  \end{center}
  \caption{A divergent subgraph (a) with terms \mbox{$\propto M^2$} and $\propto l^2$ leads to an $M^2$-independent but wrong contribution in (b) as the rest of the diagram
  produces a term $\propto \f{1}{M^2}$. This must be compensated by a mass counterterm (c).}
   \label{M2auxdivSubdia}
\end{figure}

These counterterms become important at higher orders where a divergent subgraph with a term \mbox{$\propto \f{M^2}{\eps}$} is multiplied with a term $\propto \f{1}{M^2}$
from the rest of the full diagram. Hence the wrong contribution \mbox{$\propto \f{M^2}{\eps}$} from the subgraph is not identifiable in the final result. This can be avoided by applying all possible counterterms
\mbox{$\propto M^2$} which have been computed in lower orders. Using these counterterms \mbox{$\propto M^2$} we restore the $M^2$-terms of the exact decomposition
\eqref{propdenomdecomprekursiv} order by order in perturbation theory.
An example is shown in Fig.~\ref{M2auxdivSubdia}. The one-loop diagram (a) has a divergent part $\f{l^2}{\eps}\,C_{\sss{l^2}}+\f{M^2}{\eps}\,C_{\sss{M^2}}$ if we introduce the auxiliary mass $M^2$
in the denominators of the two propagators and Taylor expand in the external momentum $l$. Only the scalar part $C_{\sss{l^2}}$ is needed in order to renormalize the Lagrangian of the (massless) theory.
The counterterm $M^2 \delta\!Z_{\sss{M^2}}^{(2\Phi)}=-\f{M^2}{\eps}\,C_{\sss{M^2}}$ however is needed at two-loop level in order to insure that only the part $\f{l^2}{\eps}\,C_{\sss{l^2}}$ of the subdiagram 
(a) of the full diagram (b) contributes to the final result.

To summarize the method, we introduce an auxiliary mass in every propagator denominator, expand in the external momenta as far as needed for the kinematic structure of the considered Greens function before
setting them to zero in the scalar part of the Greens function. Then the UV divergent part of the Greens function is computed order in order in perturbation theory by evaluating
massive Tadpole diagrams, for which we use \mbox{MATAD} \cite{MATAD}. In order to get the correct result we have to apply all possible regular UV counterterms as well as all possible counterterms
\mbox{$\propto M^2$} to the lower order diagrams.
In the Standard Model they have the form
\be
 \f{M^2}{2}\delta\!Z_{\sss{M^2}}^{(2g)}\,A_\mu^a A^{a\,\mu}{},\quad 
 \f{M^2}{2}\delta\!Z_{\sss{M^2}}^{(2W)}\,W_\mu^a W^{a\,\mu}{},\quad  
 \f{M^2}{2}\delta\!Z_{\sss{M^2}}^{(2B)}\,B_\mu B^{\mu}{},\quad
 M^2 \delta\!Z_{\sss{M^2}}^{(2\Phi)}\, \Phi^\dagger\Phi.
\label{MassenCTgauge}
\ee
Other counterterms \mbox{$\propto M^2$} and of mass dimension four can not be constructed. For fermions this is obvious, for ghosts
this is because of the kinematic structure of the ghost vertex which is always proportional to a momentum making only
kinetic counterterms \mbox{$\propto (\p_\mu\bar{c}^a)(\p^\mu c^a)$} and no term \mbox{$\propto M^2\bar{c}^ac^a$} possible.

Note, that the mass $M^2$ is introduced by hand at the level of integrands and is not a parameter of the theory. Hence these counterterms do not appear in the Lagrangian and
gauge invariance of the theory is not spoiled. The whole procedure is a mathematical trick to reduce the problem of UV renormalization counterterms to computing massive tadpoles with one scale.

The results of the calculation of the three-loop $\beta$-function for the Higgs self-interaction have been published in \cite{Chetyrkin:2012rz,Chetyrkin:2013wya}.

\section{The evolution of the Higgs self-coupling and vacuum stability} 
\begin{figure}[!b]
 \includegraphics[width=\linewidth]{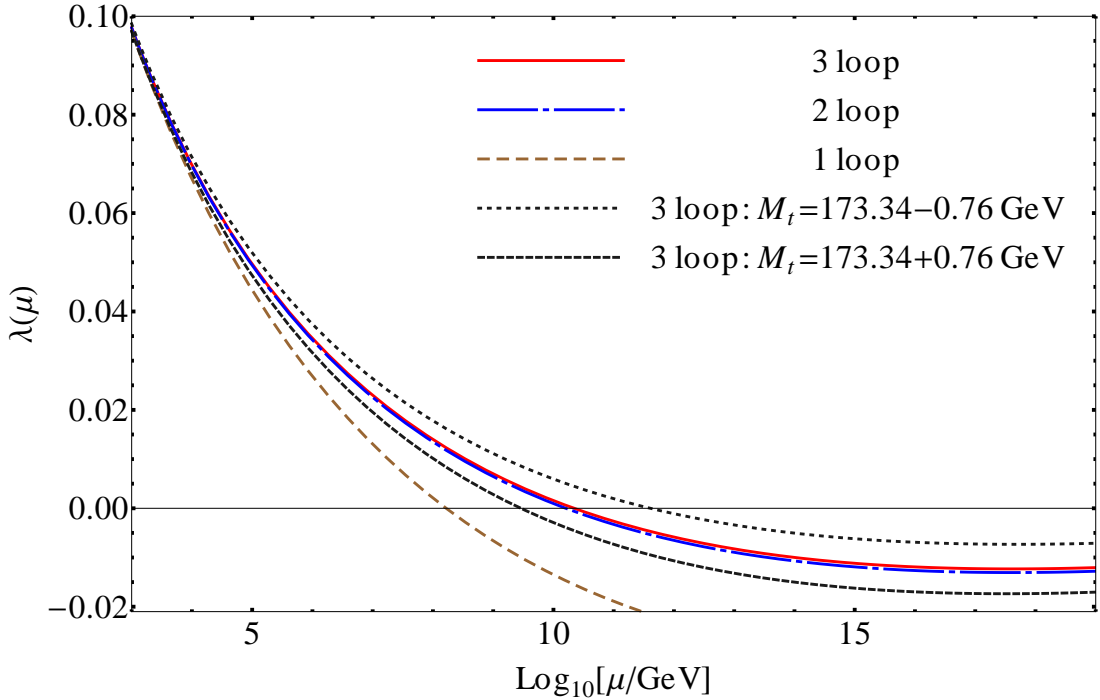}
\caption{Evolution of $\lambda$ using 1, 2 and 3 loop $\beta$-functions, top mass uncertainty} \label{lambda_evolution_Exp} 
\end{figure}
The evolution of $\lambda$ up to the Planck scale \mbox{$M_{Planck}\sim10^{19}$ GeV} 
using the $\beta$-functions for $\lambda,\yt,\gs,g_2$ and $g_1$ as well as the initial conditions from Tab.~\ref{SMKopplungenWertewitherrors}
is shown in Fig.~\ref{lambda_evolution_Exp}. 
Whereas the difference between the evolution using two-loop $\beta$-functions and the one using one-loop $\beta$-functions is significant,
the curves for three-loop $\beta$-functions and two-loop $\beta$-functions are very close. The difference between these two curves represents the theoretical uncertainty stemming from
truncating the perturbation series of the $\beta$-functions. This can be compared to the theoretical uncertainties stemming from the matching of on-shell to \msbar-parameters which are of similar size
\cite{Zoller:2013mra} and to the uncertainties of the experimental input parameters. Here the main uncertainty stems from the top mass. In Fig.~\ref{lambda_evolution_Exp} we see the three-loop evolution
of $\lambda$ for the top pole mass shifted by one \mbox{$\sigma=0.76$ GeV} \cite{ATLAS:2014wva}. The uncertainties stemming from the Higgs mass and $\als$ measurement are significantly smaller
than the one from the top mass but still larger than the theoretical ones (see e.g.~\cite{Zoller:2013mra}).

The stability of the electroweak vacuum state is a fundamental issue if the Standard Model extrapolated to high energies.
In the absence of new physics the Higgs self coupling becomes negative at \mbox{$\mathrm{Log}_{10}\lb\f{\mu}{\text{GeV}}\rb\approx 10.36$} for the best fit input parameters.
This makes a metastable scenario the most likely for the Standard Model up to the Planck scale.
Due to the calculation of three-loop $\beta$-functions for the Higgs
self-interaction and the other Standard Model couplings as well as due to improved precision in the matching relations between
on-shell and $\overline{\text{MS}}$-parameters the theoretical uncertainties are well under control.
A more precise measurement of the experimental input parameters, especially of the top mass at a linear $\text{e}^+\text{e}^-$-collider, is necessary to clarify this issue with certainty.

\section*{Acknowledgments}\vspace{-2mm}
I thank my collaborator K.~G.~Chetyrkin for invaluable discussions
and J.~H.~K\"uhn for his support and useful comments. 
This work has been supported by the Deutsche Forschungsgemeinschaft in the
Sonderforschungsbereich/Transregio SFB/TR-9 ``Computational Particle
Physics'', the Graduiertenkolleg ``Elementarteilchenphysik
bei h\"ochsten Energien und h\"ochster Pr\"azission'' 
and the ``Karlsruhe School of Elementary Particle and Astroparticle Physics: Science and Technology (KSETA)''.


\providecommand{\href}[2]{#2}\begingroup\raggedright
\endgroup

\end{document}